%% file: he3eed.tex
\documentclass[%
 reprint,
 superscriptaddress,
 showpacs,
 amsmath,amssymb,
 aps,
 prl,
]{revtex4-1}

\usepackage{graphicx} 
\usepackage{dcolumn}  
\usepackage{bm}       

\begin{document}

\title{Measurement of double-polarization asymmetries in the quasi-elastic
$^3\vec{\mathrm{He}}(\vec{\mathrm{e}},\mathrm{e}'\mathrm{d})$ process}

\input{author_list.tex}
\date{\today}

\begin{abstract}
We present a precise measurement of double-polarization asymmetries 
in the $^3\vec{\mathrm{He}}(\vec{\mathrm{e}},\mathrm{e}'\mathrm{d})$ 
reaction.  This particular process is a uniquely sensitive probe 
of hadron dynamics in $^3\mathrm{He}$ and the structure of the underlying 
electromagnetic currents.  The measurements have been performed 
in and around quasi-elastic kinematics at $Q^2 = 0.25\,(\mathrm{GeV}/c)^2$
for missing momenta up to $270\,\mathrm{MeV}/c$.  The asymmetries 
are in fair agreement with the state-of-the-art calculations in terms 
of their functional dependencies on $p_\mathrm{m}$ and $\omega$, but are
systematically offset.  Beyond the region of the quasi-elastic peak, 
the discrepancies become even more pronounced.  Thus, our measurements 
have been able to reveal deficiencies in the most sophisticated 
calculations of the three-body nuclear system, and indicate that 
further refinement in the treatment of their two- and/or three-body 
dynamics is required.
\end{abstract}


\maketitle

The $^3\mathrm{He}$ nucleus lies at the very heart of nuclear physics 
and, along with the deuteron, represents the perfect playground
to test nuclear dynamics (see, for example, \cite{Glock04,Golak2005,FBS}
and references therein).  The understanding of its structure
has far-reaching implications not only for nuclear physics itself,
but also for a variety of $^3\mathrm{He}$-based experiments seeking
to extract the neutron information by exploiting $^3\mathrm{He}$ as 
an effective neutron target.  These extractions rely on a virtually
perfect theoretical knowledge of the ground-state spin structure
of $^3\mathrm{He}$.  In particular, the statistical precision
of double-polarization experiments on $^3\mathrm{He}$ has become
comparable to the systematic uncertainty implied by our imperfect
knowledge of the polarization of the protons and the neutron within
the polarized $^3\mathrm{He}$ nucleus.  This increase in precision
needs to be matched by the best theoretical models which, in turn,
require increasingly accurate input to adjust their parameters,
complete understanding
of the spin and isospin dependence of the reaction-mechanism effects
such as final-state interactions (FSI) and meson-exchange currents (MEC),
as well as an evaluation of the possible role of three-nucleon forces.

The most fruitful approach to studying the $^3\mathrm{He}$ 
nucleus is by electron-induced knockout of protons, neutrons, 
and deuterons.  In this paper we focus on the deuteron channel.
In the $^3{\mathrm{He}}({\mathrm{e}},\mathrm{e}'\mathrm{d})$
reaction, the virtual photon emitted by the incoming electron
transfers the energy $\omega$ and momentum $\mathbf{q}$ 
to the $^3\mathrm{He}$ nucleus.  The process is best
studied by measuring its response as a function of the magnitude
of its missing momentum, which is defined as the difference between
the momentum transfer and the detected deuteron momentum,
$p_\mathrm{m} = |\,\mathbf{q} - \mathbf{p}_\mathrm{d}\,|$, hence 
$p_\mathrm{m}$ corresponds to the momentum of the recoiled proton.  

The unpolarized $^3\mathrm{He}(\mathrm{e},\mathrm{e}'\mathrm{d})$
process has been studied at Bates and NIKHEF facilities 
\cite{Keizer,Tripp,Spaltro,Spaltro02}, yielding information
on nucleon momentum distributions, isospin structure of the currents,
FSI, and MEC.  However, these measurements lacked the selective power 
of those experiments that exploit polarization.
Only a handful of such measurements exist.  The 
$^3\vec{\mathrm{He}}(\vec{\mathrm{e}},\mathrm{e}'\mathrm{p})\mathrm{pn}$ and
$^3\vec{\mathrm{He}}(\vec{\mathrm{e}},\mathrm{e}'\mathrm{p})\mathrm{d}$
channels have been studied at NIKHEF \cite{Poolman,Higinbotham:2000hc} 
and Mainz \cite{Carasco,Achenbach}, but no published data on the polarized
$^3\vec{\mathrm{He}}(\vec{\mathrm{e}},\mathrm{e}'\mathrm{d})\mathrm{p}$
exist, chiefly due to the fact that previous experiments, though
attempted, lacked present-day highly polarized beams and targets, 
which resulted in poor experimental figures-of-merit and prohibitive
uncertainties. 

It has been shown, both in the diagrammatic approach
\cite{Laget,Nagorny,NagornyS}, as well as in independent full Faddeev
calculations of the Hannover/Lisbon (H/L)
\cite{Yuan02a,Deltuva04a,Deltuva04b,Deltuva05}
and the Bochum/Krakow (B/K) \cite{GolakPRC02,GolakPRC05} groups, that the 
$^3\vec{\mathrm{He}}(\vec{\mathrm{e}},\mathrm{e}'\mathrm{d})\mathrm{p}$
reaction exhibits strong sensitivities to the sub-leading components
of the $^3\mathrm{He}$ ground-state wave-function and, possibly, 
three-nucleon forces.  Due to a particular isoscalar-isovector 
interference, this channel is also a unique source of information 
on the isospin structure of the electromagnetic current.  
It is the sensitivity brought about by the polarization degrees of freedom, 
augmented by the extended lever arm in $p_\mathrm{m}$, that lends 
the present experiment its benchmark strength.  Especially the extended 
kinematic coverage in $p_\mathrm{m}$ up to $270\,\mathrm{MeV}/c$ 
represents a crucial advantage, because the calculations enumerated 
above indicate that the manifestations of various $^3\mathrm{He}$ 
wave-function components exhibit very different signatures as functions 
of $p_\mathrm{m}$.  Moreover, these $p_\mathrm{m}$-dependencies
in each $^3\mathrm{He}$ breakup channel appear to be rather distinct.


\begin{figure}[!h]
\begin{center}
\includegraphics[width=8.0cm]{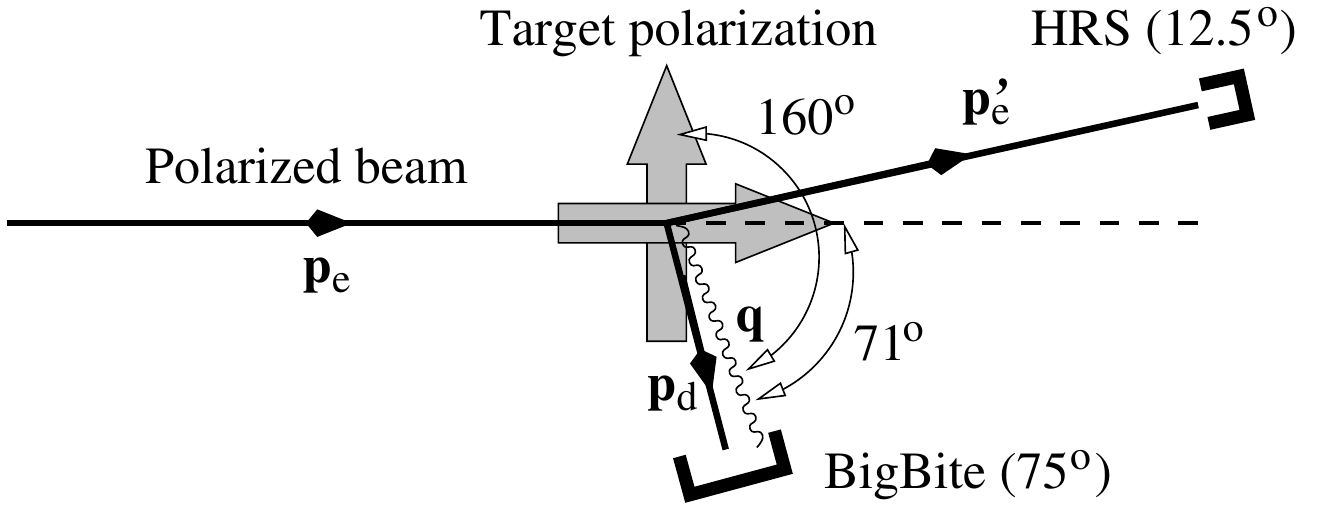}
\end{center}
\vspace*{-5mm}
\caption{Schematic drawing of the experimental setup.  
The orientations of the in-plane target polarization along
the beamline and perpendicular to it correspond to the spherical
angles about the $\vec{q}$-vector of $\theta^\ast=71^\circ$ and 
$\theta^\ast=160^\circ$, respectively, with $\phi^\ast = 0^\circ$ 
in both cases.}
\label{frame}
\end{figure}

In the case of polarized beam and polarized target, the cross-section for
the $^3\vec{\mathrm{He}}(\vec{\mathrm{e}},\mathrm{e}'\mathrm{d})\mathrm{p}$
reaction has the form
$$
\frac{\mathrm{d}\sigma(h,\vec{S})}{\mathrm{d}\Omega}
  = \frac{\mathrm{d}\sigma_0}{\mathrm{d}\Omega}
    \left[\,1+\vec{S}\cdot\vec{A}^0+h(A_\mathrm{e}+\vec{S}\cdot\vec{A})\,
    \right] \>, 
$$
where $\mathrm{d}\Omega = \mathrm{d}\Omega_\mathrm{e}\mathrm{d}E_\mathrm{e}
\mathrm{d}\Omega_\mathrm{d}$ is the differential of the phase-space
volume, $\sigma_0$ is the unpolarized cross section, 
$\vec{S}$ is the spin of the target, and $h$ is the helicity 
of the electrons.  The $\vec{A}^0$ and $A_\mathrm{e}$
are the asymmetries induced by the polarization of only the target or only 
the beam, respectively, while the spin-correlation parameter $\vec{A}$
is the asymmetry when both the beam and the target are polarized. 
If the target is polarized only in the horizontal plane defined
by the beam and scattered electron momenta (see Fig.~\ref{frame}),
the term $\vec{S}\cdot\vec{A}^0$ does not contribute \cite{Laget},
while $A_\mathrm{e}$ is parity-suppressed and is negligible
with respect to $\vec{A}$.

The orientation of the target polarization is defined by the angles
$\theta^\ast$ and $\phi^\ast$ in the frame where the $z$-axis is
along $\mathbf{q}$ and the $y$-axis is given by
$\mathbf{p}_\mathrm{e}\times\mathbf{p}_\mathrm{e}'$.
Any component of $\vec{A}$, i.~e.~the asymmetry at given 
$\theta^\ast$ and $\phi^\ast$ is then 
\begin{equation}
A(\theta^\ast,\phi^\ast) =
 \frac{(\mathrm{d}\sigma/\mathrm{d}\Omega)_+ -
       (\mathrm{d}\sigma/\mathrm{d}\Omega)_-}
      {(\mathrm{d}\sigma/\mathrm{d}\Omega)_+ + 
       (\mathrm{d}\sigma/\mathrm{d}\Omega)_-} \>,
\label{A}
\end{equation}
where the subscript signs represent the beam helicities.  
In this paper we report measurements of these asymmetries 
in the $^3\vec{\mathrm{He}}(\vec{\mathrm{e}},\mathrm{e}'\mathrm{d})$
process in quasi-elastic kinematics at the average four-momentum 
transfer of $Q^2 = \mathbf{q}^2 - \omega^2 = 0.25\,(\mathrm{GeV}/c)^2$,
performed during the E05-102 experiment at the Thomas Jefferson 
National Accelerator Facility in experimental Hall~A \cite{HallANIM}.


We used an electron beam with an energy of $2.425\,\mathrm{GeV}$ and
currents in excess of $10\,\mu\mathrm{A}$.  The beam was longitudinally
polarized, with an average polarization of $P_\mathrm{e} = (84.3\pm 2.0)\,\%$
measured by a M{\o}ller polarimeter.  The beam helicity was flipped at 
$30\,\mathrm{Hz}$ in $+--+$ or $-++-$ structures in pseudorandom sequence.


The beam was incident on a $40\,\mathrm{cm}$-long glass target cell
containing the $^3\mathrm{He}$ gas at approximately $9.3\,\mathrm{bar}$
(corresponding to the surface-area density of $0.043\,\mathrm{g/cm}^2$),
which was polarized by hybrid spin-exchange optical pumping
\cite{Walker,Appelt,Babcock,Singh}.  The in-plane orientation 
of the polarization was maintained by two pairs of Helmholtz coils.  
To measure the $A(160^\circ,0^\circ)$ asymmetry, the coils were used 
to rotate the $^3\mathrm{He}$ spin vector to the left of the beam, 
at $160^\circ$ with respect to $\mathbf{q}$, 
while for the $A(71^\circ,0^\circ)$ asymmetry it was maintained 
along the beamline, at $71^\circ$ with respect to $\mathbf{q}$ 
(both values are averages over the whole spread of angles).  
Electron paramagnetic resonance and nuclear magnetic resonance 
\cite{Abragam,Romalis,Babcock2}
were used to monitor the polarization of the target, $P_\mathrm{t}$, 
which was between $50\,\%$ and $60\,\%$ throughout the experiment 
and was taken into account on a run-by-run basis.


The scattered electrons were detected by a High-Resolution
magnetic Spectrometer (HRS) positioned at
$\theta_\mathrm{e}=12.5^\circ$ and equipped with a detector package
consisting of a pair of scintillator planes used for triggering
and time-of-flight measurements, vertical drift chambers
for particle tracking, and a gas \v{C}erenkov counter and lead-glass 
calorimeters for particle identification.


The ejected deuterons and protons were detected by the large-acceptance
spectrometer BigBite equipped with a detector package optimized for hadron
detection \cite{BBNIM}, consisting of a pair of multi-wire drift chambers
used for tracking and two scintillator planes used for triggering,
time-of-flight determination, and particle identification.


The electrons in the HRS were selected by applying cuts on the 
\v{C}erenkov detector signals.  The most reliable selection of deuterons
in BigBite was achieved by using graphical cuts in two-dimensional
histograms of scintillator ADC (particle energy loss) vs.~particle
momentum as determined from track reconstruction.  Depending on
the kinematics, $(1-2)\,\%$ of protons may become misidentified
as deuterons, which has a minute influence on the final results.
For the extraction of the asymmetries, only electron-deuteron coincidence
events were retained, based on the measurement of coincidence time.
Additional cuts on the location of the target vertex (to eliminate
the contribution from the cell walls) and on the quality of 
the reconstructed tracks were used to further purify the event sample.


The experimental asymmetry for each orientation of the target polarization
was determined as the relative difference between the number of coincidence
events (after all cuts and background subtraction) corresponding
to positive and negative beam helicities,
$A_\mathrm{exp} = (N_+ - N_-) / (N_+ + N_-)$, where $N_+$ and $N_-$
have been corrected for helicity-gated beam charge asymmetry, 
dead time and radiative effects.  The corresponding physics asymmetries 
were calculated as $A = A_\mathrm{exp} / (P_\mathrm{e}P_\mathrm{t})$.

\begin{figure}[hbtp]
\begin{center}
\includegraphics[width=8.5cm]{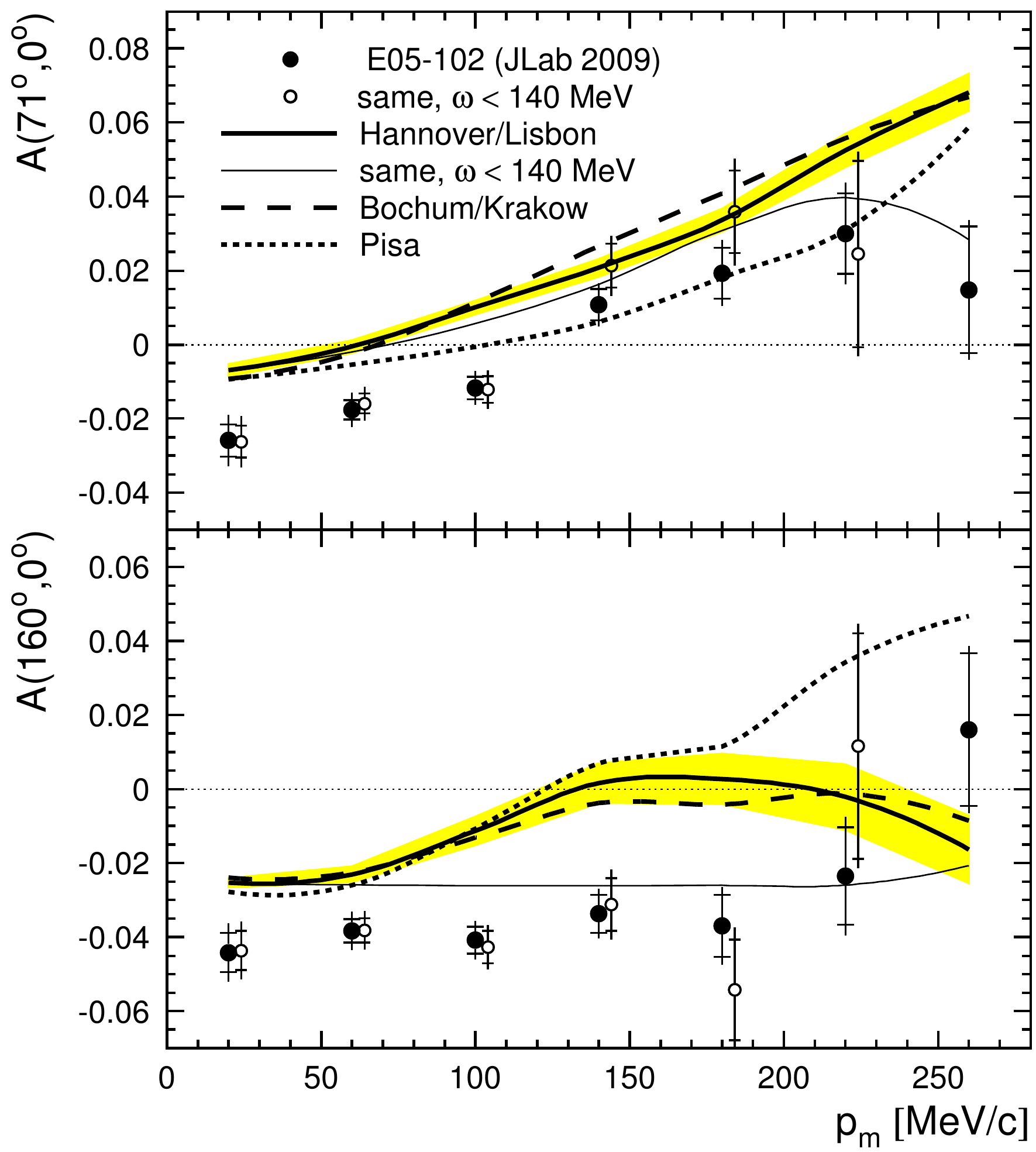}
\end{center}
\vspace*{-5mm}
\caption{(Color online.) The asymmetries $A(71^\circ,0^\circ)$ (top)
and $A(160^\circ,0^\circ)$ (bottom) in the quasi-elastic
$^3\vec{\mathrm{He}}(\vec{\mathrm{e}},\mathrm{e}'\mathrm{d})$ process
as functions of missing momentum, compared to the acceptance-averaged
calculations of the Hannover/Lisbon, Bochum/Krakow and Pisa groups.
The double error bars on the data denote the statistical and total
uncertainties (statistical and systematical part added linearly).
The shaded (yellow) bands indicate the uncertainties implied by 
the acceptance-averaging procedure.  They have been placed
on the H/L curves because that calculation has been
averaged over the finest mesh.  Empty symbols (shifted for clarity) 
and thin curves denote the data and the corresponding H/L calculation 
with a cut on the quasi-elastic peak.}
\label{altpm}
\end{figure}

The asymmetries as functions of $p_\mathrm{m}$ are shown in Fig.~\ref{altpm}.
The largest contribution to the systematic error comes from the relative uncertainty 
of the target polarization, $P_\mathrm{t}$, which has been estimated at $5\,\%$, 
followed by the uncertainty due to protons contaminating the deuteron sample, 
an effect that translates into a systematic error of $3\,\%$ at low hadron momenta 
to less than $1\,\%$ at high momenta.  The absolute error of the beam polarization, 
$P_\mathrm{e}$, was $2\,\%$, while the error due to the uncertainty of the target 
orientation angle $\theta^*$ was $0.6\,\%$.  Due to finite spectrometer acceptances 
there was a spread in $\theta^*$ and $\phi^*$ around their nominal values, 
which has been taken into account in the theory acceptance-averaging procedure.
The total systematic uncertainty (all items added in quadrature) is $7\,\%$ (relative).  
The resolution in $p_\mathrm{m}$ is driven mostly by the relative momentum resolution 
of BigBite, which is $\approx 2\,\%$ for all momenta \cite{BBNIM}, 
while the contribution of the $\omega$ resolution ($2.2\cdot 10^{-4}$) is negligible.  
Hence, the kinematic dependence of the systematic uncertainties is very small, 
and the possible smearing of the asymmetries has been excluded.

Figure~\ref{altpm} also shows the results of the 
state-of-the-art three-body calculations of the Hannover/Lisbon 
\cite{Yuan02a,Deltuva04a,Deltuva04b,Deltuva05}, Bochum/Krakow 
\cite{GolakPRC02,GolakPRC05} and Pisa \cite{marcucci05} groups.
The B/K calculations are based on the AV18 nucleon-nucleon
potential \cite{av18,av18p} and involve a complete treatment 
of FSI and MEC, but do not include three-nucleon forces; 
the Coulomb interaction is taken into account in the $^3\mathrm{He}$ 
bound state.  The H/L calculations are based on 
the coupled-channel extension of the charge-dependent Bonn potential 
\cite{cdbonn} and also include FSI and MEC, while the $\Delta$ isobar 
is added as an active degree of freedom providing a mechanism 
for an effective three-nucleon force and for exchange currents.  
Point Coulomb interaction is added in the partial waves involving 
two charged baryons.  The Pisa calculations are based on the AV18 
interaction model (augmented by the Urbana IX three-nucleon force
\cite{uix}), in which full inclusion of FSI is taken into account by means 
of the variational pair-correlated hyper-spherical harmonic expansion, 
as well as MEC.  Coulomb interaction is included in full (not only
in the $^3\mathrm{He}$ ground state).  In contrast to the B/K
and H/L approaches, the Pisa calculations are not genuine 
Faddeev calculations but are of equivalent precision and are expected 
to account for all relevant reaction mechanisms.  


Due to the extended momentum and angular acceptances of HRS 
and BigBite, the theoretical asymmetries were averaged over 
these acceptances.  The averaging was performed over the whole
accepted region of the $(E_\mathrm{e}',\theta_\mathrm{e})$ plane 
in $63$ bins for the H/L calculations and $35$ bins for 
the B/K and Pisa calculations.  In each of these bins, 
the asymmetries were evaluated on a mesh of $p_\mathrm{m}$ 
and deuteron azimuthal angles with respect to $\mathbf{q}$,
and interpolated.  The acceptance-averaged 
theoretical asymmetries for each $p_\mathrm{m}$ bin and their errors 
originating in this procedure were then obtained by evaluating 
a weighted average and mapped onto the seven $p_\mathrm{m}$-bins
used to display the measured asymmetries.


Neither of the three considered theories exactly reproduces the measured 
$A(160^\circ,0^\circ)$ asymmetry --- which is fairly constant at about 
$-4\,\%$ throughout the $p_\mathrm{m}$ range --- except when a quasi-elastic cut 
($\omega < 140\,\mathrm{MeV}$) is applied.  The improved agreement is not surprising 
as all present calculations are known to perform better in the region of 
the quasi-elastic peak, while their reliability is expected to deteriorate
in the dip region and beyond due to the opening of the pion production threshold
and increasing influences of resonances, all of which have so far not been
taken into account.  A hint of the zero-crossing 
of the measured asymmetry at high $p_\mathrm{m}$ appears to be
mirrored by the theoretical one, but it occurs at much lower $p_\mathrm{m}$,
and the predicted asymmetries, in addition to exhibiting a mismatch
in the functional form, are roughly a factor of two too small.
On the other hand, the measured $A(71^\circ,0^\circ)$ shows a clear 
zero-crossing at $p_\mathrm{m} \approx 130\,\mathrm{MeV}/c$ seen also 
in all three calculations, although it again occurs at much lower 
$p_\mathrm{m}$.


\begin{figure}[!h]
\begin{center}
\includegraphics[width=6.0cm]{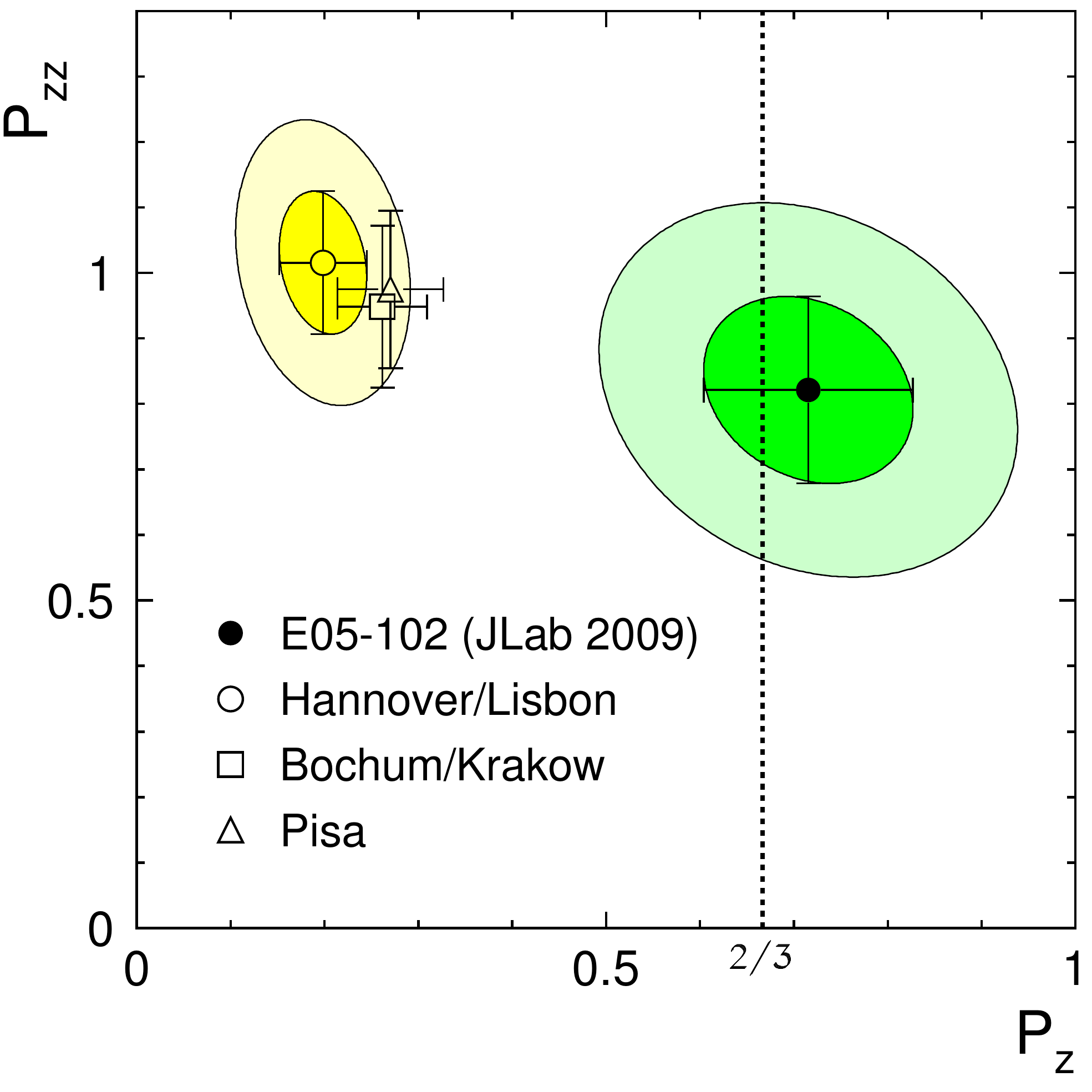}
\end{center}
\vspace*{-5mm}
\caption{(Color online.) Effective vector and tensor deuteron 
polarizations (spin orientations) $P_z$ and $P_{zz}$ in 
$^3\mathrm{He}$ extracted from the data and theoretical predictions 
at $p_\mathrm{m}\to 0$ in the approximation of e-d elastic 
scattering (with $1\sigma$ and $2\sigma$ covariance ellipses
on the experiment (green) and the numerically most reliable theory 
interpolation (H/L, yellow)).  If the spin part of the $^3\mathrm{He}$ 
wave-function were given simply by a Clebsch-Gordan combination 
of the $\mathrm{p}$ and $\mathrm{d}$ parts, one would expect 
$P_z = 2/3$ and $P_{zz} = 0$.}
\label{pzpzz}
\end{figure}

One could argue that, at low $p_\mathrm{m}$, the asymmetries 
for the deuteron knock-out process
$^3\vec{\mathrm{He}}(\vec{\mathrm{e}},\mathrm{e}'\mathrm{d})$
should be similar to the asymmetries for electron scattering almost
elastically off a polarized deuteron within polarized $^3\mathrm{He}$.  
To assess this instructive, if simplistic, view we have equated
the measured $A(160^\circ,0^\circ)$ and $A(71^\circ,0^\circ)$
for $^3\mathrm{He}$ at low $p_\mathrm{m}$ 
($p_\mathrm{m} \le 40\,\mathrm{MeV}/c$) with the corresponding 
$\vec{\mathrm{e}}$-$\vec{\mathrm{d}}$ asymmetries, 
computed at the same $(\theta^\ast,\phi^\ast)$ and $Q^2$.
By using appropriate deuteron form-factors \cite{Abbott}
one can extract the vector and tensor spin orientations 
of the deuteron, $P_z$ and $P_{zz}$, inside $^3\mathrm{He}$.  
From the data, 
we obtain $P_z(\mathrm{exp}) = 0.72\pm 0.11$ 
and $P_{zz}(\mathrm{exp}) = 0.82\pm 0.14$, indicating 
--- under the above assumptions --- that the deuteron in $^3\mathrm{He}$ 
is strongly polarized (spin ``up''), the third component of zero 
being disfavored due to $P_{zz} \approx 1$.  By applying the same 
procedure on the theoretical predictions, we obtain $P_z$ 
of $0.20$--$0.27$ and $P_{zz}$ of $0.95$--$1.01$, depending 
on the model (see Fig.~3).   
In this approximation, the incomplete theoretical description of
both $^3\vec{\mathrm{He}}(\vec{\mathrm{e}},\mathrm{e}'\mathrm{d})$
asymmetries at low $p_\mathrm{m}$ maps to an inadequacy in just
one parameter, $P_z$, which is underestimated by a factor
of about $3$, while $P_{zz}$ is only slightly overestimated.

\begin{figure}[h]
\begin{center}
\includegraphics[width=8.5cm]{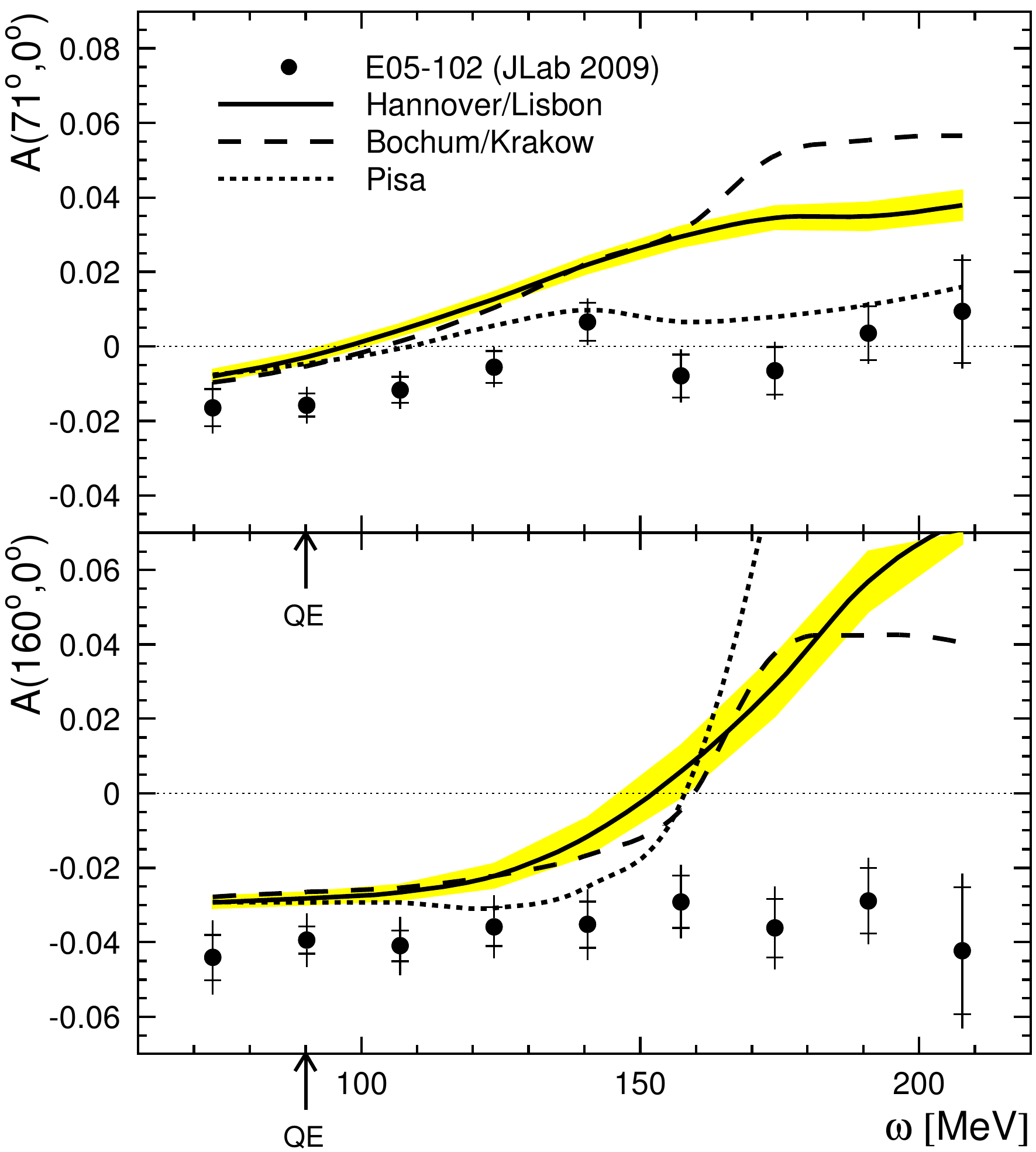}
\end{center}
\vspace*{-5mm}
\caption{The asymmetries $A(71^\circ,0^\circ)$ 
and $A(160^\circ,0^\circ)$ in the 
$^3\vec{\mathrm{He}}(\vec{\mathrm{e}},\mathrm{e}'\mathrm{d})$ process
as functions of energy transfer.  The arrows denote the approximate
location of the quasi-elastic peak.}
\label{altom}
\end{figure}

The asymmetries as functions of energy transfer, $\omega$, are shown 
in Fig.~\ref{altom}.  At low $\omega$, both measured asymmetries,
$A(160^\circ,0^\circ)$ and $A(71^\circ,0^\circ)$, are fairly well 
reproduced in all approaches in terms of 
the functional form, but not in magnitude: again, there is 
a systematic offset of the asymmetries of about one or two percent 
(absolute).  At high $\omega$, all calculated asymmetries deviate 
from the measured ones, even in the H/L prediction that 
has been evaluated on the finest mesh, indicating that the dynamic input
in the theoretical treatment of the process in the dip region is incomplete.  

In conclusion, we have provided the world-first, high-precision data
on a high-level physics observable at two different spin settings 
in a broad kinematic range.  Three most sophisticated theoretical 
treatments of the $^3\mathrm{He}$ system are able to qualitatively 
account for the bulk of our data set; given the small magnitude 
of the asymmetries and the subtle interplay of the myriad
of their ingredients, the agreement is actually quite good 
--- in spite of the systematic offsets in $p_\mathrm{m}$- 
and $\omega$-dependencies and deviations occurring in the dip region.
Up to the level of this agreement, the basic theoretical assumptions
on the hadron dynamics and on the structure of the electromagnetic
currents have been justified, and it appears that a consistent 
$^3\mathrm{He}$ ground-state wave-function has been employed.  
However, the large precision of our measurements 
has been able to reveal deficiencies in the calculations, indicating
a need for further refinement in the treatment of their two- and/or
three-body dynamics.  In fact, the detailed anatomy of the $p_\mathrm{m}$ 
dependence of asymmetries is already the subject of a major ongoing 
theoretical effort.  Among other things, our data will now allow one 
to check which leading and sub-leading components make up 
the employed $^3\mathrm{He}$ wave-function that are consistent with 
the assumed dynamics, and thereby significantly advance our knowledge
of the three-nucleon system.

\begin{acknowledgments}
We thank the Jefferson Lab Hall A and Accelerator Operations technical 
staff for their outstanding support.  This material is based upon work 
supported by the U.S. Department of Energy, Office of Science, 
Office of Nuclear Physics under contract DE-AC05-06OR23177.  
This work was supported in part by the Polish National Science Center 
under Grant No.~DEC-2013/10/M/ST2/00420.  The numerical calculations 
of the Bochum/Krakow group were partly performed on the supercomputer 
cluster of the JSC, J\"ulich, Germany.
\end{acknowledgments}

\bibliographystyle{apsrev4-1}
\bibliography{he3eed}

\end{document}

%% file: author_list.tex
\author{M.~Mihovilovi\v{c}} \email[Presently at Institut f\"ur Kernphysik, Johannes-Gu\-ten\-berg-Universit\"at, Mainz, Germany]{} \affiliation{Jo\v{z}ef Stefan Institute, SI-1000 Ljubljana, Slovenia}
\author{G.~Jin} \affiliation{University of Virginia, Charlottesville, VA 22908, USA}
\author{E.~Long} \affiliation{Kent State University, Kent, OH 44242, USA} 
\author{Y.-W.~Zhang} \affiliation{Rutgers University, New Brunswick, NJ 08901, USA}
\author{K.~Allada} \affiliation{Thomas Jefferson National Accelerator Facility, Newport News, VA 23606, USA}
\author{B.~Anderson} \affiliation{Kent State University, Kent, OH 44242, USA}
\author{J.~R.~M.~Annand} \affiliation{Glasgow University, Glasgow G12 8QQ, Scotland, United Kingdom}
\author{T.~Averett} \affiliation{The College of William and Mary, Williamsburg, VA 23187, USA}
\author{W.~Boeglin} \affiliation{Florida International University, Miami, FL 33181, USA}
\author{P.~Bradshaw} \affiliation{The College of William and Mary, Williamsburg, VA 23187, USA}
\author{A.~Camsonne} \affiliation{Thomas Jefferson National Accelerator Facility, Newport News, VA 23606, USA}
\author{M.~Canan} \affiliation{Old Dominion University, Norfolk, VA 23529, USA}
\author{G.~D.~Cates} \affiliation{University of Virginia, Charlottesville, VA 22908, USA}
\author{C.~Chen} \affiliation{Hampton University, Hampton, VA 23669, USA}
\author{J.~P.~Chen} \affiliation{Thomas Jefferson National Accelerator Facility, Newport News, VA 23606, USA}
\author{E.~Chudakov} \affiliation{Thomas Jefferson National Accelerator Facility, Newport News, VA 23606, USA}
\author{R.~De~Leo} \affiliation{Universit\`a degli studi di Bari Aldo Moro, I-70121 Bari, Italy}
\author{X.~Deng} \affiliation{University of Virginia, Charlottesville, VA 22908, USA}
\author{A.~Deltuva} \affiliation{Center for Nuclear Physics, University of Lisbon, P-1649-003 Lisbon, Portugal} \affiliation{Institute for Theoretical Physics and Astronomy, Vilnius University, LT-01108 Vilnius, Lithuania}
\author{A.~Deur} \affiliation{Thomas Jefferson National Accelerator Facility, Newport News, VA 23606, USA}
\author{C.~Dutta} \affiliation{University of Kentucky, Lexington, KY 40506, USA}
\author{L.~El~Fassi} \affiliation{Rutgers University, New Brunswick, NJ 08901, USA}
\author{D.~Flay} \affiliation{Temple University, Philadelphia, PA 19122, USA}
\author{S.~Frullani} \affiliation{Istituto Nazionale Di Fisica Nucleare, INFN/Sanita, Roma, Italy}
\author{F.~Garibaldi} \affiliation{Istituto Nazionale Di Fisica Nucleare, INFN/Sanita, Roma, Italy}
\author{H.~Gao} \affiliation{Duke University, Durham, NC 27708, USA}
\author{S.~Gilad} \affiliation{Massachusetts Institute of Technology, Cambridge, MA 02139, USA}
\author{R.~Gilman} \affiliation{Rutgers University, New Brunswick, NJ 08901, USA}
\author{O.~Glamazdin} \affiliation{Kharkov Institute of Physics and Technology, Kharkov 61108, Ukraine}
\author{J.~Golak} \affiliation{M. Smoluchowski Institute of Physics, Jagiellonian University, PL-30059 Krak\'ow, Poland}
\author{S.~Golge} \affiliation{Old Dominion University, Norfolk, VA 23529, USA}
\author{J.~Gomez} \affiliation{Thomas Jefferson National Accelerator Facility, Newport News, VA 23606, USA}
\author{O.~Hansen} \affiliation{Thomas Jefferson National Accelerator Facility, Newport News, VA 23606, USA}
\author{D.~W.~Higinbotham} \affiliation{Thomas Jefferson National Accelerator Facility, Newport News, VA 23606, USA}
\author{T.~Holmstrom} \affiliation{Longwood College, Farmville, VA 23909, USA}
\author{J.~Huang} \affiliation{Massachusetts Institute of Technology, Cambridge, MA 02139, USA}
\author{H.~Ibrahim} \affiliation{Cairo University, Cairo, Giza 12613, Egypt}
\author{C.~W.~de~Jager} \affiliation{Thomas Jefferson National Accelerator Facility, Newport News, VA 23606, USA}
\author{E.~Jensen} \affiliation{Christopher Newport University, Newport News VA 23606, USA}
\author{X.~Jiang} \affiliation{Los Alamos National Laboratory, Los Alamos, NM 87545, USA}
\author{M.~Jones} \affiliation{Thomas Jefferson National Accelerator Facility, Newport News, VA 23606, USA}
\author{H.~Kang} \affiliation{Seoul National University, Seoul, Korea}
\author{J.~Katich} \affiliation{The College of William and Mary, Williamsburg, VA 23187, USA}
\author{H.~P.~Khanal} \affiliation{Florida International University, Miami, FL 33181, USA}
\author{A.~Kievsky} \affiliation{INFN-Pisa, I-56127 Pisa, Italy}
\author{P.~King} \affiliation{Ohio University, Athens, OH 45701, USA}
\author{W.~Korsch} \affiliation{University of Kentucky, Lexington, KY 40506, USA}
\author{J.~LeRose} \affiliation{Thomas Jefferson National Accelerator Facility, Newport News, VA 23606, USA}
\author{R.~Lindgren} \affiliation{University of Virginia, Charlottesville, VA 22908, USA}
\author{H.-J.~Lu} \affiliation{Huangshan University, People's Republic of China}
\author{W.~Luo} \affiliation{Lanzhou University, Lanzhou, Gansu, 730000, People's Republic of China}
\author{L.~E.~Marcucci} \affiliation{Physics Department, Pisa University, I-56127 Pisa, Italy}
\author{P.~Markowitz} \affiliation{Florida International University, Miami, FL 33181, USA}
\author{M.~Meziane} \affiliation{The College of William and Mary, Williamsburg, VA 23187, USA}
\author{R.~Michaels} \affiliation{Thomas Jefferson National Accelerator Facility, Newport News, VA 23606, USA}
\author{B.~Moffit} \affiliation{Thomas Jefferson National Accelerator Facility, Newport News, VA 23606, USA}
\author{P.~Monaghan} \affiliation{Hampton University, Hampton, VA 23669, USA}
\author{N.~Muangma} \affiliation{Massachusetts Institute of Technology, Cambridge, MA 02139, USA}
\author{S.~Nanda} \affiliation{Thomas Jefferson National Accelerator Facility, Newport News, VA 23606, USA}
\author{B.~E.~Norum} \affiliation{University of Virginia, Charlottesville, VA 22908, USA}
\author{K.~Pan} \affiliation{Massachusetts Institute of Technology, Cambridge, MA 02139, USA}
\author{D.~Parno} \affiliation{Carnegie Mellon University, Pittsburgh, PA 15213, USA}
\author{E.~Piasetzky} \affiliation{Tel Aviv University, Tel Aviv 69978, Israel}
\author{M.~Posik} \affiliation{Temple University, Philadelphia, PA 19122, USA}
\author{V.~Punjabi} \affiliation{Norfolk State University, Norfolk, VA 23504, USA}
\author{A.~J.~R.~Puckett} \affiliation{Los Alamos National Laboratory, Los Alamos, NM 87545, USA}
\author{X.~Qian} \affiliation{Duke University, Durham, NC 27708, USA}
\author{Y.~Qiang} \affiliation{Thomas Jefferson National Accelerator Facility, Newport News, VA 23606, USA}
\author{X.~Qui} \affiliation{Lanzhou University, Lanzhou, Gansu, 730000, People's Republic of China}
\author{S.~Riordan} \affiliation{University of Virginia, Charlottesville, VA 22908, USA}
\author{A.~Saha} \email[Deceased.]{} \affiliation{Thomas Jefferson National Accelerator Facility, Newport News, VA 23606, USA}
\author{P.~U.~Sauer} \affiliation{Institute for Theoretical Physics, University of Hannover, D-30167 Hannover, Germany}
\author{B.~Sawatzky} \affiliation{Thomas Jefferson National Accelerator Facility, Newport News, VA 23606, USA}
\author{R.~Schiavilla} \affiliation{Thomas Jefferson National Accelerator Facility, Newport News, VA 23606, USA} \affiliation{Old Dominion University, Norfolk, VA 23529, USA}
\author{B.~Schoenrock} \affiliation{Northern Michigan University, Marquette, MI 49855, USA}
\author{M.~Shabestari} \affiliation{University of Virginia, Charlottesville, VA 22908, USA}
\author{A.~Shahinyan} \affiliation{Yerevan Physics Institute, Yerevan, Armenia}
\author{S.~\v{S}irca} \email[Corresponding author: ]{simon.sirca@fmf.uni-lj.si} \affiliation{University of Ljubljana, SI-1000 Ljubljana, Slovenia}\affiliation{Jo\v{z}ef Stefan Institute, SI-1000 Ljubljana, Slovenia}
\author{R.~Skibi\'nski} \affiliation{M. Smoluchowski Institute of Physics, Jagiellonian University, PL-30059 Krak\'ow, Poland}
\author{J.~St.~John} \affiliation{Longwood College, Farmville, VA 23909, USA}
\author{R.~Subedi} \affiliation{George Washington University, Washington, D.C. 20052, USA}
\author{V.~Sulkosky} \affiliation{Massachusetts Institute of Technology, Cambridge, MA 02139, USA}
\author{W.~A.~Tobias} \affiliation{University of Virginia, Charlottesville, VA 22908, USA}
\author{W.~Tireman} \affiliation{Northern Michigan University, Marquette, MI 49855, USA}
\author{G.~M.~Urciuoli} \affiliation{Istituto Nazionale Di Fisica Nucleare, INFN/Sanita, Roma, Italy}
\author{M.~Viviani} \affiliation{INFN-Pisa, I-56127 Pisa, Italy}
\author{D.~Wang} \affiliation{University of Virginia, Charlottesville, VA 22908, USA}
\author{K.~Wang} \affiliation{University of Virginia, Charlottesville, VA 22908, USA}
\author{Y.~Wang} \affiliation{University of Illinois at Urbana-Champaign, Urbana, IL 61801, USA}
\author{J.~Watson} \affiliation{Thomas Jefferson National Accelerator Facility, Newport News, VA 23606, USA}
\author{B.~Wojtsekhowski} \affiliation{Thomas Jefferson National Accelerator Facility, Newport News, VA 23606, USA}
\author{H.~Wita{\l}a} \affiliation{M. Smoluchowski Institute of Physics, Jagiellonian University, PL-30059 Krak\'ow, Poland}
\author{Z.~Ye} \affiliation{Hampton University, Hampton, VA 23669, USA}
\author{X.~Zhan} \affiliation{Massachusetts Institute of Technology, Cambridge, MA 02139, USA}
\author{Y.~Zhang} \affiliation{Lanzhou University, Lanzhou, Gansu, 730000, People's Republic of China}
\author{X.~Zheng} \affiliation{University of Virginia, Charlottesville, VA 22908, USA}
\author{B.~Zhao} \affiliation{The College of William and Mary, Williamsburg, VA 23187, USA}
\author{L.~Zhu} \affiliation{Hampton University, Hampton, VA 23669, USA}
\collaboration{The Jefferson Lab Hall A Collaboration} \noaffiliation